\documentclass[nofootinbib,superscriptaddress,twocolumn,aps,prc,showpacs,10pt]{revtex4-1}

\usepackage{bm}
\usepackage{amsmath}
\usepackage{amssymb}
\usepackage{graphicx}
\usepackage{mathrsfs}
\usepackage{hyperref}
\usepackage{color}

\newcommand{\red}[1]{\textcolor{red}{{#1}}}
\newcommand{\blue}[1]{\textcolor{blue}{{#1}}}

\begin{document}

\title{$\Xi$ hypernuclei $^{15}_{\Xi}$C, $^{12}_{\Xi}$Be and $\Xi N$ two-body interaction}

\author{Yusuke Tanimura}
\affiliation{Department of Physics, Tohoku University, Sendai, 980-8578, Japan}
\affiliation{Graduate Program on Physics for the Universe, Tohoku University, Sendai, 980-8578, Japan}

\author{Hiroyuki Sagawa} 
\affiliation{Center for Mathematics and Physics, the Univeristy of Aizu, Aizu-Wakamatsu, Fukushima 965-8580, Japan}
\affiliation{RIKEN, Nishina Center, Wako, Saitama 351-0198, Japan}

\author{Tingting Sun}
\affiliation{School of Physics and Microelectronics, Zhengzhou University, Zhengzhou 450001, China}

\author{Emiko Hiyama}
\affiliation{Department of Physics, Tohoku University, Sendai, 980-8578, Japan}
\affiliation{RIKEN, Nishina Center, Wako, Saitama 351-0198, Japan}

\date{\today}

\begin{abstract}
We study the energy spectra of $\Xi$ hypernuclei $^{15}_{\Xi}$C and $^{12}_{\Xi}$Be 
with a relativistic mean field (RMF) model with meson exchange $\Xi N$ interactions. 
The RMF parameters are optimized to reproduce the average energy  of KINKA and IRRAWADDY events for the ground state and also the average energy of KISO and IBUKI events for  the excited state in $^{15}_{\Xi}$C. 
The potential depth of average $\Xi N$ mean field potential is found to be about $-12$ MeV in the nuclear matter limit. 
We further introduce the two-body $s$- and $p$-wave interactions between valence nucleons and $\Xi$ particle. 
We found that the $s$-wave interaction deduced from the HAL LQCD results is rather weak to obtain the energy difference between IRRAWADDY and KINKA events. 
The $p$-wave interaction is added and fitted to reproduce the energy difference. 
The resulting interaction together with the $s$-wave one gives a reasonable energy 
simultanously for the IBUKI event as an excited $\Xi_p$ state. 
The model is further applied to predict the energy spectrum of $^{12}_{\Xi}$Be. 
\end{abstract}

\maketitle

\section{Introduction}
The study of  interaction of the strangeness $S=-2$ sector is important in
hypernuclear physics.  To this end,  so far several $YY$ potential models such as Njimegen group \cite{esc} and chiral effective field theory \cite{chiral} have been proposed.     
However, due to the difficulty of performing hyperon($Y$)-hyperon($Y$) 
scattering experiment, the proposed potentials have 
a large ambiguity depending on the models.   Instead, 
as an alternative way to obtain information on $S=-2$ sector, more pragmatic models were proposed to reproduce  empirical data 
with less ambiguity.  
For such approaches, it is crucial to study structure of double strangeness 
hypernuclei such as double $\Lambda$ hypernuclei and $\Xi$ hypernuclei. 
Experimentally, some double $\Lambda$ hypernuclei, $^6_{\Lambda\Lambda}$He (NAGARA event)
\cite{NAGARA,DEMACHI},
$^{10}_{\Lambda \Lambda}$Be (DEMACHI-YANAGI event)\cite{DEMACHI}, 
and $^{11}_{\Lambda \Lambda}$Be (HIDA \cite{DEMACHI} and MIKAGE events
 \cite{ekawa}) 
have been observed by KEK-E373 and J-PARC-E07 experiments. 
In comparisons with the data and theoretical calculations,
it was found that the  $^1S_0$ of $\Lambda \Lambda$ interaction is attractive, but  much less attractive than that of $\Lambda N$ interaction.
Further analysis of J-PARC-E07 experiment is still in progress.

Regarding $\Xi$ hypernuclei,  the first evidence of bound $\Xi$ hypernucelus,
$^{15}_{\Xi}$C ($^{14}{\rm N}+\Xi^-$), was reported in  2015 by KEK-E373 experiment, 
which is named KISO event \cite{kiso}.
At that time, there were two possible $\Xi^-$ binding energy interpretations 
from the analyses of the kinematic formation and decay modes,
$\Xi^-+^{14}{\rm N} \rightarrow  ^{15}_{\Xi}{\rm C} \rightarrow 
^{10}_{\Lambda}{\rm Be}
+^5_{\Lambda}$He. 
One is $B_{\Xi}\equiv-E(^{14}{\rm N})=4.38 \pm 0.25$ MeV  assuming 
the ground state of $^{10}_{\Lambda}$Be after the decay from $^{15}_{\Xi}$C.
The other is $B_{\Xi}=1.11 \pm 0.25$ MeV  assuming the decay from 
the first excited state of
$^{10}_{\Lambda}$Be.
Afterwards, due to the revision of the observed binding energy of 
the $^{10}_{\Lambda}$Be, the $B_{\Xi}$ of KISO event was updated accordingly to be $3.87 \pm 0.21$ MeV or $1.03 \pm 0.18$ MeV \cite{HiNa18}.
Thanks to the observation of KISO event, it was found that $\Xi$-nucleus interaction 
should be attractive,  and it was requested to interpret theoretically 
that KISO event is the observation of either  $^{14}{\rm N(g.s.)}+\Xi (1s)$ or
$^{14}{\rm N(g.s.)}+\Xi (1p)$. 
Motivated by the observed data in Ref.\cite{kiso},
some of the present authors (T.T. S, H. S and E. H.) studied the
structure of $^{15}_{\Xi}$C using  relativistic mean field (RMF) and 
Skyrme-Hartree-Fock (SHF) models, and interpreted KISO event as $^{14}{\rm N(g.s.)}+\Xi (1p)$ \cite{Sun16}.
Afterwards, $^{14}{\rm N(g.s.)}+\Xi (1p)$ has been observed as a IBUKI event with  the binding energy $B_{\Xi}=1.27 \pm 0.21$ MeV \cite{ibuki}.
Very recently, two  new  events  $^{14}{\rm N(g.s.)}+\Xi(1s)$
 has been reported  by J-PARC-E07.
One is called KINKA event and the other is called IRRAWADDY event \cite{irra}.
The observed $B_{\Xi}$ of the former is $8.00 \pm 0.77$ MeV or
$4.96 \pm 0.77$ MeV, and that of the latter is $6.27 \pm 0.27$ MeV.  
These observations could give a great contribution to the
information on $s$-wave $\Xi N$ interaction. For the further study of
$\Xi N$ interaction, it is planned to perform a search experiment of
$^{12}_{\Xi}$Be at J-PARC by $^{12}{\rm C}(K^-,K^+)$ reaction. 

Considering this situation, it is important to reproduce 
all of observed events theoretically.
We consider that KINKA and IRRAWADDY events are partner of
the ground-state doublet and that IBUKI and KISO events are also
partner of the excited-state doublet 
because there should exists  $\sigma \cdot \sigma$ and 
$\tau \cdot \tau$ dependent terms in the  $\Xi N$ interaction, whereas
spin-orbit contribution is negligible for the ground state doublet of 
$^{15}_{\Xi}$C. In this paper, we adopt the RMF model with 
the meson-exchange  $\Xi N$ interactions, which
were also employed in the previous paper \cite{Sun16}.
In addition, in this paper, we introduce the $\sigma \cdot \sigma$ and 
$\tau \cdot \tau$ dependent two-body $\Xi N$ interactions
 for valence nucleons and $\Xi$ particle.
Parameters of the two-body $\Xi N$ interaction 
 should be optimized so as to reproduce the
IRRAWADDY event and Kiso event.
Moreover, the $\Xi N$ interaction employed in $^{15}_{\Xi}$C,
is applied to study of  $^{12}_{\Xi}$Be.
It is planned to produce  $^{12}_{\Xi}$Be by $(K^-,K^+)$ reaction 
using $^{12}$C target as J-PARC-E70 experiment.
To encourage the experiment, we predict energy spectra of this hypernucleus.

Since the core nuclei, $^{14}$N and $^{11}$B, have the 
compact shell structure,  it is hardly compressed by the addition of
$\Xi$ hyperon.  Thus, 
the RMF approach is suited for the present work.
In fact, in the previous paper \cite{Sun16} by RMF, we accomplished  
interpreting KISO event as $^{14}{\rm N(g.s.)}+\Xi (1p)$.
Thus  it is highly expected that we can interpret the IRRAWADDY event and predict
low-lying level structure of $^{12}_{\Xi}$Be in a consistent manner.

This paper is organized as follows. Sec. \ref{sec:model} is devoted for the introduction 
of the model.
Results of $^{15}_\Xi$C and $^{12}_\Xi$Be are presented in Sec. \ref{sec:results}. 
The average $\Xi N$ mean-field potentials for various $\Xi$ nuclei is also discussed in Sec. \ref{sec:results}.
Summary is given in Sec. \ref{sec:summary}.

\section{Model}
\label{sec:model}

\subsection{Relativistic mean-field model for $\Xi$ hypernuclei}


To describe the $\Xi$ hypernucleus, we employ a meson-exchange model 
with non-linear couplings for RMF theory \cite{BB77}. 
The Lagrangian density is given by 
\begin{eqnarray}
{\cal L} &=& 
\bar\psi_N(i\partial\!\!\!/-m_N)\psi_N
+\bar\psi_\Xi(i\partial\!\!\!/-m_\Xi)\psi_\Xi
\nonumber \\
&&
+\frac{1}{2}(\partial_\mu\sigma)(\partial^\mu\sigma) - \frac{1}{2}m_\sigma^2\sigma^2
-\frac{c_3}{3}\sigma^3-\frac{c_4}{4}\sigma^4
\nonumber \\
&&
-\frac{1}{4}G^{\mu\nu}G_{\mu\nu} + \frac{1}{2}m_\omega^2\omega^\mu\omega_\mu
+\frac{d_4}{4}(\omega^\mu\omega_\mu)^2
\nonumber \\
&&
-\frac{1}{4}\vec R^{\mu\nu}\cdot\vec R_{\mu\nu}
+ \frac{1}{2}m_\rho^2\vec\rho^\mu\cdot\vec\rho_\mu
\nonumber \\
&&
 -\frac{1}{4}F^{\mu\nu}F_{\mu\nu}
\nonumber \\
&&
-\bar\psi_N\left(g_{\sigma N}\sigma+ g_{\omega N}\omega\!\!\!/ 
+g_{\rho N}\vec\rho\!\!\!/\cdot\vec\tau_N 
+eA\!\!\!/\frac{1-\tau_{N,3}}{2}
\right)\psi_N
\nonumber \\
&&
-\bar\psi_\Xi\left(g_{\sigma\Xi}\sigma
+ g_{\omega\Xi}\omega\!\!\!/ 
+ g_{\rho\Xi}\vec\rho\!\!\!/ \cdot\vec\tau_\Xi
\right.
\nonumber\\
&&\hspace{1cm}
\left.
+\frac{f_{\omega\Xi}}{4m_\Xi}G_{\mu\nu}\sigma^{\mu\nu}
+eA\!\!\!/\frac{-1-\tau_{\Xi,3}}{2}
\right)\psi_\Xi, 
\label{eq:L}
\end{eqnarray}
where $\psi_N$ and $\psi_\Xi$ are nucleon and $\Xi$ hyperon fields, and 
$G^{\mu\nu}=\partial^\mu \omega^\nu- \partial^\nu  \omega^\mu$, 
$\vec R^{\mu\nu}=\partial^\mu \vec\rho^\nu- \partial^\nu\vec\rho^\mu$, and
$F^{\mu\nu}=\partial^\mu A^\nu- \partial^\nu A^\mu$ 
are the field tensors of the vector mesons $\omega$ and $\rho$, and the photon, respectively. 
$\vec\tau_N$ and $\tau_\Xi$ are the Pauli matrix in the isospin space. 
We adopt PK1 parameter set \cite{pk1} for the nucleon-meson couplings. 
The $\rho-\Xi$ coupling constant is taken to be $g_{\rho\Xi} = g_{\rho N}$ as 
in Refs. \cite{MJ94,Sun16}. 
The remaining coupling constants $g_{\sigma\Xi}$, $g_{\omega\Xi}$, and $f_{\omega\Xi}$  
in the hyperon sector are fitted to experimental data (IRRAWADDY \cite{irra} and 
KINKA events, see Sec. \ref{sec:results}).
The $\Xi$ hyperon mass $m_\Xi$ is taken to be $1321.7$ MeV \cite{pdg20}. 

The model Lagrangian given above is solved within the mean-field and the no-sea approximations. 
We impose the time-reversal invariance and charge conservation of the mean-field state, 
{\it i.e.}, the time-odd or charged vector fields vanish.  
The only non-zero components are their time-like and neutral components, 
$\omega^0$, $\rho^0_3$, and $A^0$, 
where the subscript $3$ on the $\rho$ meson field means the third component in the isospace. 

The scalar-isoscalar density $\rho_S$, vector-isoscalar and -isovector densities $j^0$ and $j_3^0$ 
of nucleon are defined in terms of the single-particle wave functions of nucleon $\psi_k^{(N)}$ as 
\begin{eqnarray}
\rho_S(\bm r) &=& \sum_{k\in{\rm occ}}\bar\psi_k^{(N)}(\bm r)\psi_k^{(N)}(\bm r),
\label{eq:rhos}
\\
j^0(\bm r) &=& \sum_{k\in{\rm occ}}\bar\psi_k^{(N)}(\bm r)\gamma^0\psi_k^{(N)}(\bm r), 
\\
j^0_3(\bm r) &=& \sum_{k\in{\rm occ}}\bar\psi_k^{(N)}(\bm r)\gamma^0\tau_3\psi_k^{(N)}(\bm r), 
\end{eqnarray}
where $k$ runs over the occupied nucleon states. 
The scalar, vector, and tensor densities of $\Xi$ hyperon are defined in terms of the single-particle 
wave function of $\Xi$ hyperon $\psi_k^{(\Xi)}$ as
\begin{eqnarray}
\rho_{S\Xi}(\bm r) &=& \bar\psi_k^{(\Xi)}(\bm r)\psi_k^{(\Xi)}(\bm r),
\\
j^0_\Xi(\bm r) &=& \bar\psi_k^{(\Xi)}(\bm r)\gamma^0\psi_k^{(\Xi)}(\bm r), 
\\
j^0_{\Xi3}(\bm r) &=& \bar\psi_k^{(\Xi)}(\bm r)\gamma^0\tau_3\psi_k^{(\Xi)}(\bm r), 
\\
\bm V_{T\Xi}(\bm r) &=&
\bar\psi_k^{(\Xi)}(\bm r)i\bm\alpha\psi_k^{(\Xi)}(\bm r), 
\end{eqnarray}
where $k$ here denotes the occupied $\Xi$ hyperon state. 

The equations of motion for the meson and electromagnetic fields read 
\begin{eqnarray}
(-\bm\nabla^2+m_\sigma^2)\sigma &=& -g_{\sigma N}\rho_S-c_3\sigma^2 - c_4\sigma^3
\nonumber \\
&&
-g_{\sigma\Xi}\rho_{S\Xi}, 
\\
(-\bm\nabla^2+m_\omega^2)\omega^0 &=&
 g_{\omega N} j^0 - d_4\left(\omega^0\right)^3
\nonumber \\
&&
+g_{\omega\Xi}j^0_\Xi
+\frac{f_{\omega\Xi}}{2m_\Xi}\bm\nabla\cdot\bm V_{T\Xi}, 
\\
(-\bm\nabla^2+m_\rho^2)\rho^0_3  &=& g_{\rho N} j^0_3 + g_{\rho \Xi} j^0_{\Xi3},
\\
-\bm\nabla^2A^0 &=& e\left[\frac{1}{2}(j^0-j^0_3)-\frac{1}{2}(j^0_\Xi-j^0_{\Xi3})\right]. 
\end{eqnarray}

The Dirac equation for the nucleon single-particle wave function is given by 
\begin{equation}
\left[-i\bm\alpha\cdot\bm\nabla + V_N + \beta(m_N+S_N)\right]\psi_k^{(N)}
=\epsilon_k\psi_k^{(N)}, 
\end{equation}
where 
\begin{eqnarray}
S_N &=& g_{\sigma N}\sigma, \\
V_N &=& g_{\omega N}\omega^0 + g_{\rho N}\rho^0_3\tau_{N,3} + eA^0\frac{-1-\tau_{N,3}}{2}. 
\end{eqnarray}
Note that $\tau_{N,3}=+1$ for neutron and $\tau_{N,3}=-1$ for proton. 
The Dirac equation for the single-particle wave function of $\Xi$ hyperon is given by 
\begin{equation}
\left[-i\bm\alpha\cdot\bm\nabla + V_\Xi + \beta(m_\Xi+S_\Xi)
+i\beta\bm\alpha\cdot\bm T_\Xi \right]\psi_k^{(\Xi)}
=\epsilon_k\psi_k^{(\Xi)}, 
\end{equation}
where 
\begin{eqnarray}
S_\Xi &=& g_{\sigma \Xi}\sigma, \\
V_\Xi &=& g_{\omega \Xi}\omega^0 + g_{\rho\Xi}\rho^0_3\tau_{\Xi,3} + eA^0\frac{1+\tau_{\Xi,3}}{2},
\label{eq:V_Xi} \\ 
\bm T_\Xi &=& \frac{f_{\omega\Xi}}{2m_\Xi}
\left(-\bm\nabla\omega^0\right). 
\label{eq:potl}
\end{eqnarray}

The set of equations, Eqs. (\ref{eq:rhos}-\ref{eq:potl}), are solved self-consistently. 
In the numerical calculations, we further impose the spherical symmetry, and 
the radial coordinate is discritized up to 40 fm with the step size of $0.1$ fm. 
The spurious $\Xi-\Xi$ interaction effect via the $\rho$ meson is removed 
in the same way as in Ref. \cite{MJ94}. 

\begin{widetext}
\subsection{Residual interactions}

We consider spin-isospin-dependent residual interaction, which will be introduced 
in Sec. \ref{sec:results}. 
Once the residual interaction is given, 
its effect on the energy spectra of $^{15}_{\Xi}$C and $^{12}_{\Xi}$Be 
will be estimated by the first-order perturbation theory. 
We construct the unperturbed states based on the RMF wave functions. 

The $^{14}$N subsystem of $^{15}_{\Xi}$C nucleus is described as the $p_{1/2}$ neutron and proton coupled 
to the total spin-parity $J^{\pi_{np}}_{np}=1^+$ and isospin $T_{np} = 0$ on top of the inert $^{12}$C core. 
The $\Xi$ particle is then coupled to the nucleon pair to make the total quantum numbers $J^\pi T$, 
where $J^\pi$ is the total spin-parity of the system, and $T=1/2$ in the case of $^{15}_{\Xi}$C. 
The energy shift due to the residual interaction $V_{N\Xi}$ is given as 
\begin{eqnarray}
\Delta E
&=&
\left\langle \left[\left[\nu 1p_{1/2}~\pi 1p_{1/2}\right]^{1^+}\Xi nl_j\right]^{J^\pi} \right|V_{N\Xi}
\left| \left[\left[\nu 1p_{1/2}~\pi 1p_{1/2}\right]^{1^+}\Xi nl_j\right]^{J^\pi} \right\rangle, 
\end{eqnarray}
where we take here the proton-neutron formalism, and the obvious isospin quantum number $T=1/2$ is implicit. 
$nlj$ denotes the radial quantum number, orbital and total angular momentum of the $\Xi$ particle, respectively. 

The $^{12}_\Xi$Be nucleus is regarded as a particle-hole state, {\it i.e.}, a $p_{3/2}$ nucleon hole 
and a 1$s_{1/2}$ $\Xi$ particle on top of the $^{12}$C core, which are coupled to good angular momentum and isospin $J^\pi T$. 
In the case of $^{12}_{\Xi}$Be, the energy shift due to the residual interaction is given as 
\begin{eqnarray}
\Delta E 
&=&
\left\langle \left[N 1p_{3/2}^{-1}~\Xi nl_j\right]^{J^\pi T} \right|V_{N\Xi}
\left| \left[N 1p_{3/2}^{-1}~\Xi nl_j\right]^{J^\pi T} \right\rangle. 
\end{eqnarray}

\end{widetext}




\section{Results}\label{sec:results}

\subsection{$\Xi N$ interaction and energy spectra of $^{15}_{\Xi}$C}

We adjust the parameters, $g_{\sigma\Xi}$, $g_{\omega\Xi}$, and $f_{\omega\Xi}$ 
so as to reproduce the observed $B_{\Xi}=6.27 \pm 0.24$ MeV, IRRAWADDY event for $^{14}{\rm N(g.s.)}+\Xi(1s)$, and
 $B_{\Xi}=1.03 \pm 0.18$ MeV, KISO event for $^{14}{\rm N(g.s.)}+\Xi(1p)$.
It should be noted that we have other two observed data,
IBUKI event, $B_{\Xi}=1.27 \pm 0.21$ MeV, KINKA event 
$B_{\Xi}=8.00 \pm 0.77$ MeV or $4.96 \pm 0.77$ MeV, which are
expected to be $^{14}{\rm N(g.s.)}+\Xi(1p)$ and
$^{14}{\rm N(g.s.)}+\Xi(1s)$, respectively.
First, assuming the KINKA and IRRAWADDY events are the ground-state spin doublet, 
and KISO and IBUKI events are a part of the excited-state multiplets, 
we fit the $\Xi-\omega$ Yukawa couplings $g_{\sigma\Xi}$ and $g_{\omega \Xi}$
so that $B_{\Xi(1s)}$ is located between IRRAWADDY and KINKA, and 
spin-averaged $B_{\Xi(1p)}$ is close to KISO and IBUKI. 
Furthermore, we introduce $\Xi-\omega$ tensor coupling $f_{\omega\Xi}$ to produce
spin-orbit force of $\Xi N$ so that $B_{\Xi(1p_{3/2})}$ and $B_{\Xi(1p1/2)}$ are 0.86 MeV and 0.72 MeV, respectively.
The parameters thus determined leads to $B_{\Xi(1s)}=5.69$ MeV for the 
RMF ground-state binding energy. 
The resultant values of the coupling constants are tabulated in Table \ref{tb:x-cc}. 
In view of the order of magnitude, our phenomenological values are consistent with 
the values with the naive quark counting and the quark model \cite{CW91}. 
Note that $g_{\omega\Xi}/g_{\omega N}\approx 0.5$ is close to the value adopted in 
the recent quark mean field model study \cite{Hu21}. 

\begin{table}
\caption{$\Xi$-meson coupling constants. The ``Quark model'' values $1/3$ for $g_{\sigma\Xi}/g_{\sigma N}$ in the first row 
and for $g_{\omega\Xi}/g_{\omega N}$ in the second row come from the naive quark counting, 
and $-0.4$ for $f_{\omega\Xi}/g_{\omega\Xi}$ from a quark model in Ref. \cite{CW91}. }
\begin{tabular}{ccc}
\hline\hline
Coupling & Fitted & Quark model \\
\hline
$g_{\sigma\Xi}/g_{\sigma N}$ & $0.497$    & $1/3$\\
$g_{\omega\Xi}/g_{\omega N}$ & $0.5377$ & $1/3$\\
$f_{\omega\Xi}/g_{\omega\Xi}$ & $-0.8$  & $-0.4$\\
\hline\hline
\end{tabular}
\label{tb:x-cc}
\end{table}

Next, we study spin-isospin-dependent residual $\Xi N$ interactions. 
The $s$-wave interaction is defined by
\begin{eqnarray} \label{eq:Vres_s}
V_{N\Xi}&=&
\sum_{i\in{\rm nucleons}} (v_\sigma\bm\Sigma_i\cdot\bm\Sigma_\Xi
+v_\tau\vec\tau_i\cdot\vec\tau_\Xi     \nonumber \\
&+&v_{\sigma\tau}\bm\Sigma_i\cdot\bm\Sigma_\Xi\vec\tau_i\cdot\vec\tau_\Xi)\delta(\bm r_i-\bm r_\Xi), 
\label{eq:V_NXs}
\end{eqnarray}
where $\bm\Sigma = \left(\begin{array}{cc}
\bm\sigma & 0 \\
0 & \bm\sigma
\end{array}\right)$ is the spin operator acting on a Dirac spinor. 
In this work, we use the information on a potential based on first principle 
lattice QCD simulations, the HAL QCD $\Xi N$ potential (HAL potential) \cite{hal20}
in the following way:
First we calculate volumes of HAL potential defined by
\begin{eqnarray}
V_{\beta}=\int d^3r f_\beta(r) V_{\beta}^{\rm HAL}(r),
\end{eqnarray} 
where $\beta=0, \sigma, \tau, \sigma \tau$, and
\begin{eqnarray}
f_\beta(r)=\frac{1}{1+{\rm exp}[-(r-R_\beta)/d]}.
\end{eqnarray} 
Here, we take $d=0.1$ fm and $R_\beta=(0.234, 0.816, 0.548, 0.684)$ fm for 
the spin-isospin channels $(2S+1, 2T+1)=(11, 31, 13, 33)$, respectively, 
so that $V_{\beta}^{\rm HAL}(R)=0$. As a result, the factors $f_\beta(r)$ suppress
the repulsive cores of the HAL potential, which would induce the short-range correlation 
missing in our simple framework. 
Also, we take zero-range reduction of RMF meson-exchange interaction
 \cite{TH12,BMMR02}: 
\begin{eqnarray}
\alpha &=& -\frac{g_{\sigma N}g_{\sigma \Xi}}{m_\sigma^2}
 + \frac{g_{\omega N}g_{\omega \Xi}}{m_\omega^2}
 = -327\ {\rm MeV\ fm}^3. 
\end{eqnarray}
Thus, finally, we employ
\begin{eqnarray}
v_\sigma &=& (V_\sigma/V_0) \alpha = 4.87\ {\rm MeV\ fm}^3, \label{eq33}
\\
v_{\tau} &=& (V_\tau/V_0) \alpha = 26.83\ {\rm MeV\ fm}^3, \label{eq34}
\\
v_{\sigma\tau} &=& (V_{\sigma\tau}/V_0) \alpha = -53.56\ {\rm MeV\ fm}^3. \label{eq35}
\end{eqnarray}

Among the spin and isospin dependent terms of $\Xi N$ interaction given in Eq. \eqref{eq:V_NXs}, 
only $v_\sigma$ contributes to
the binding energy of  $^{15}_{\Xi}$C since the core is  the isospin $T=0$ nucleus,
$^{14}$N. As seen in Eq. \eqref{eq33},  the value $v_\sigma$ is rather small so that we cannot
reproduce the spin-doublet state of  IRRAWADDY and KINKA events by the $s$-wave interaction. 
It should also be noted that the HAL potential provides the $s$-wave $\Xi N$ interaction only,
and a missing part of $\Xi N$ interaction, 
$p$-wave spin and isospin dependent terms, might be important in $^{15}_{\Xi}$C since the core $^{14}$N is a $p$-shell nucleus.

Next, we introduce a $p$-wave spin dependent interaction defined by
\begin{eqnarray}  \label{p-wave}
V^p_{N\Xi}&=&
\sum_{i\in{\rm nucleons}} v_\sigma^p\bm\Sigma_i\cdot\bm\Sigma_\Xi
\overleftarrow{\bm\nabla}\cdot
\delta(\bm r_i-\bm r_\Xi)\overrightarrow{\bm\nabla}, 
\end{eqnarray}
where  $v_\sigma^p$ is parameter to be optimized by the  IRRAWADDY and KINKA events.

\begin{figure}
\includegraphics[width=\linewidth]{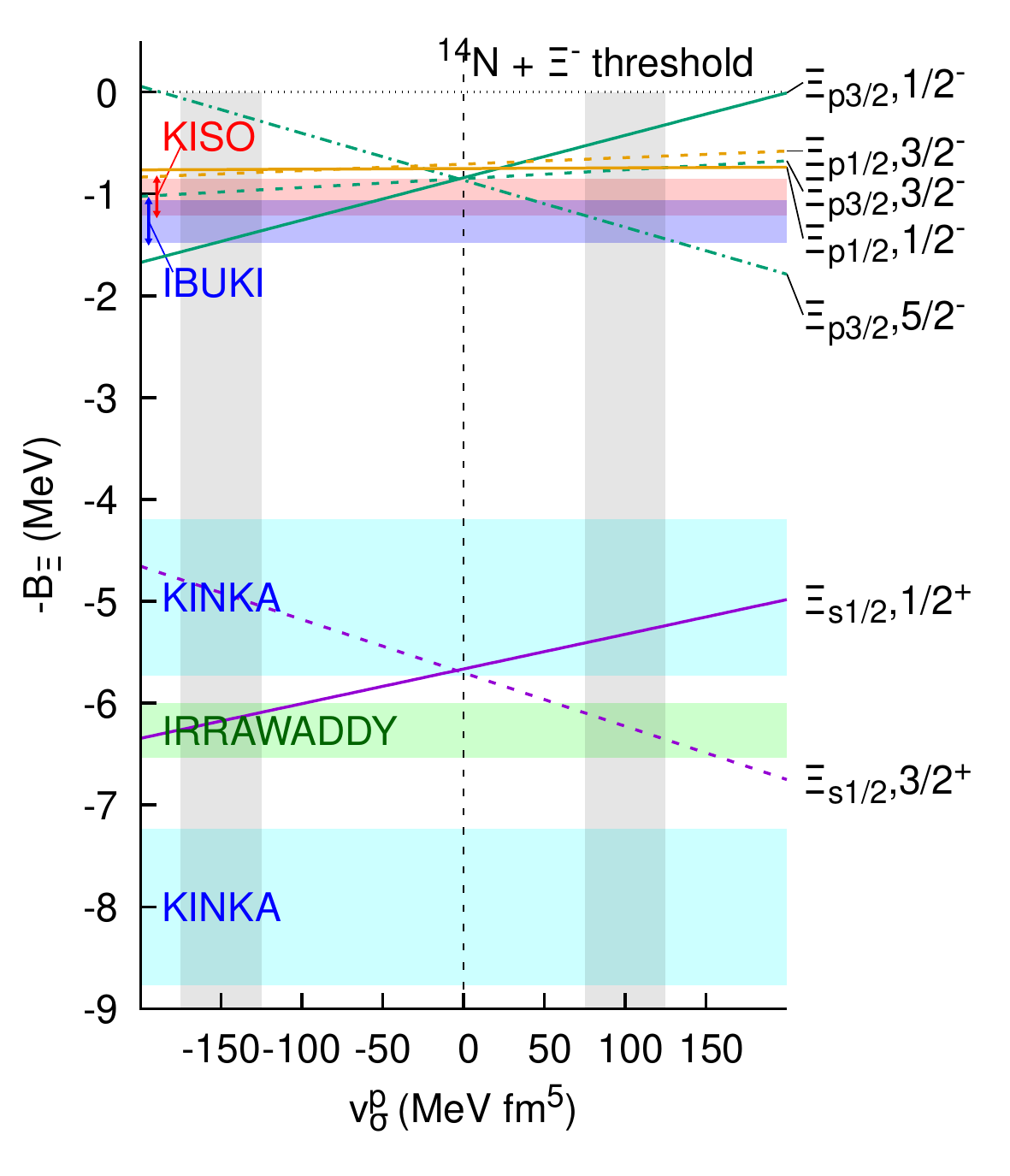}
\caption{The estimated energy spectrum of $^{15}_{\Xi}$C nucleus 
as a function of the $p$-wave spin-spin interaction strength $v_{\sigma}^p$. 
Shown by lines are the energies of the levels labeled by the orbital occupied by  
the $\Xi^-$ particle, $\Xi_{lj}$, and the total spin parity, $J^\pi$. 
The experimental data of $B_{\Xi^-}$ are shown by color bands. 
The measured binding energy of KISO event was reported in Refs. \cite{kiso,HiNa18}, 
IRRAWADDY and KINKA events in Ref. \cite{irra}, and IBUKI event in Ref. \cite{ibuki}. 
For KISO and KINKA events there are two possibilities each that correspond to 
the ground and an excited state of the decay products.}
\label{fig:spectrum15}
\end{figure}

Fig. \ref{fig:spectrum15} shows the energy spectrum of $^{15}_{\Xi}$C as a function of
$p$-wave spin-spin interaction $v_\sigma^p$.
We see that for the negative value of $v_\sigma^p \sim -150$ MeV fm$^5$,
the energies of $J=1/2^+$ and $3/2^+$ are in good agreement with IRRAWADDY event
and KINKA events, respectively.
In this case, IRRAWADDY event (KINKA event) can be interpret as observation of the
ground (excited) state.
On the other hand, for the positive value of $v_\sigma^p \sim 100$ MeVfm$^5$
the energies of $J=3/2^+$ and $1/2^+$ are  in good agreement with IRRAWADDY event
and KINKA events, respectively.
In this case, IRRAWADDY event (KINKA event) can be interpret as observation of the
ground (excited) state, too. 
It should be noticed that the opitimized $p$-wave $N\Xi$ interaction is more or less the same magnitude of the $p$-wave $NN$ ineteraction in the Skyrme EDF, which was determined by using the empirical nuclear structure information of several double-closed shell nucleus \cite{VB72}.

\begin{figure}
\includegraphics[width=\linewidth]{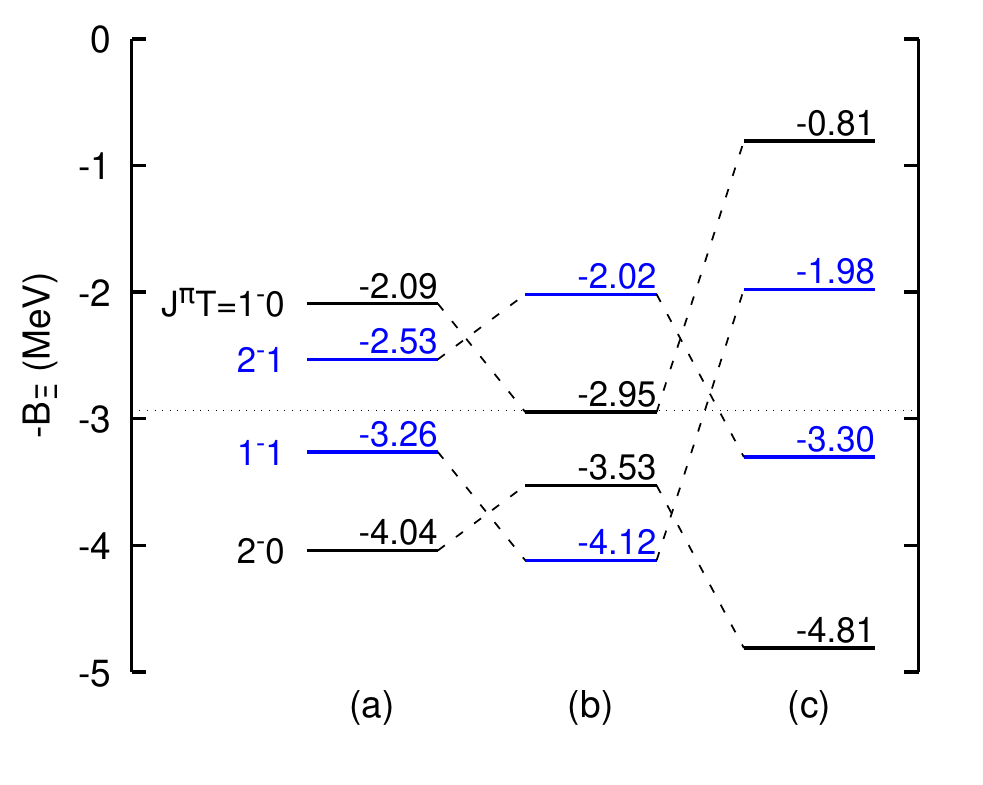}
\caption{The estimated energy spectra of $^{12}_{\Xi}$Be nucleus, 
in which $\Xi^-$ occupies the $1s$ state.  (a) 
shows the spectrum calculated by RMF with the $s$-wave interaction as given in Eq. \eqref{eq:Vres_s}. 
(b) and (c) 
represent the spectra  calculated by RMF 
with the $p$-wave interaction as given in Eq. \eqref{p-wave} together with  the $s$-wave. 
The spectrum (b) is obtained with a repulsive $p$-wave interaction $v_\sigma^p=100$ MeV fm$^5$, 
while  the spectrum  (c) is with an attractive $p$-wave one $v_\sigma^p=-150$ MeV fm$^5$. 
The four states obtained with RMF  without the residual interactions are degenerate and the energy is shown by the straight dotted line.
The energies of the levels are  shown in unit of MeV. }
\label{fig:spectrum12}
\end{figure}

\subsection{Energy spectra of $^{12}_{\Xi}$Be}
Fig. \ref{fig:spectrum12} shows the estimated energy spectra of $^{12}_{\Xi}$Be nucleus, 
in which $\Xi^-$ occupies the $1s$ state.  Fig. \ref{fig:spectrum12}(a) 
shows the spectrum calculated by RMF with only the $s$-wave interaction as given 
in Eq. \eqref{eq:Vres_s}. 
Fig. \ref{fig:spectrum12}(b) and \ref{fig:spectrum12}(c) represent the spectra calculated 
with the $p$-wave interaction as given in Eq. \eqref{p-wave} as well as the $s$-wave interaction. 
The spectrum (b) is obtained with a repulsive $p$-wave interaction $v_\sigma^p=100$ MeV fm$^5$, 
while the spectrum (c) is with an attractive one $v_\sigma^p=-150$ MeV fm$^5$. 
The four states by RMF results without the residual interactions are degenerate and the energy is shown by the straight dotted line.

In Fig.2 (a), we illustrate the spectra calculated with the HAL potential.
We see that the ground state of  $^{12}_{\Xi}$Be becomes $T=0, J^\pi=2^-$.
Qualitatively, the spectrum in Fig. 2(a) with the $s$-wave interaction can be understood in the following argument. 
Since the $\sigma\tau$ term is dominant in the $s$-wave interaction, we demonstrate the $s$-wave effect to calculate the two-body $N\Xi$ matrix element taking only the
spin-isospin operators, 
\begin{eqnarray}  \label{eq38}
&&\Delta E 
\propto  \nonumber \\
&&\left\langle \left[N1p_{3/2}^{-1}~\Xi nl_j\right]^{J^\pi T} \right|\bm\sigma_N\cdot\bm\sigma_\Xi\vec\tau_N\cdot\vec\tau_\Xi
\left| \left[N1p_{3/2}^{-1}~\Xi nl_j\right]^{J^\pi T} \right\rangle. \nonumber \\
\end{eqnarray} 
The isospin part is estimated to be
\begin{eqnarray}   \label{eq39}
\left\langle T|\vec\tau_N\cdot\vec\tau_\Xi|T \right\rangle= 
 \begin{cases}
        1 & \text{for} \,\,\, T=1, \\
         -3  & \text{for} \,\,\, T=0.
         \end{cases}
 \end{eqnarray}   
 The spin part of  Eq. \eqref{eq38} is also calculated as the reduced matrix form  
 \begin{eqnarray}  \label{eq40}
&&  \left\langle \left[\nu 1p_{3/2}^{-1}~\Xi nlj\right]^{J^\pi T} \right\|\bm\sigma_N\cdot\bm\sigma_\Xi
\left\| \left[\nu 1p_{3/2}^{-1}~\Xi nlj\right]^{J^\pi T} \right\rangle \nonumber \\
&=&(-)^{(J+1)}\left\{\begin{array}{lcl}
1/2 & 3/2 & J \\
 3/2 & 1/2  & 1 \\
\end{array}\right\}
\nonumber \\
&\times& 
\langle p_{3/2}\|\bm \sigma\|p_{3/2}\rangle  \langle s_{1/2}\|\bm\sigma_\Xi\|s_{1/2}\rangle, 
\end{eqnarray}
where we use the $6j$ symbol, which is given by
\begin{equation}
(-)^{(J+1)}\left\{\begin{array}{lcl}
1/2 & 3/2 & J \\
 3/2 & 1/2  & 1 \\
\end{array}\right\}=
 \begin{cases}
        -\left(\frac{1}{40}\right)^{1/2}   & \text{for} \,\,\, J=2, \\
       +\left(\frac{5}{24}\right)^{1/2}    & \text{for} \,\,\, J=1.
         \end{cases}
         \end{equation}
For the spin-isospin channel, the strength $v_{\sigma \tau}$ is attractive so that $J=2, T=0$ state obtains the most attractive energy, while 
$J=1, T=0$ receives the most repulsive effect in Fig. 2(a).  For the $T=1$ state, the spin-isospin interaction works in opposite directions because of Eq. \eqref{eq39}, which  makes smaller contributions for $J=2, T=1$ and $J=1, T=1$ states, respectively, compared with 
$T=0$ states. Mathematical structures of $p$-wave interactions are more complicated, but essentially follows the features of $s$-wave interactions for the attractive $p$-wave in Fig. 2(c), where both the $s$- and $p$-wave interactions work coherently.
For the repulsive $p$-wave case in Fig. 2(b), 
two interactions cancel partially  for $T=0$ states, 
but the net effect gives more energy shift for $T=1$ 
states than $T=0$ states.

\subsection{$\Xi$ mean-field potential and $\Xi$ single-particle energy}

\begin{figure}
\includegraphics[width=\linewidth]{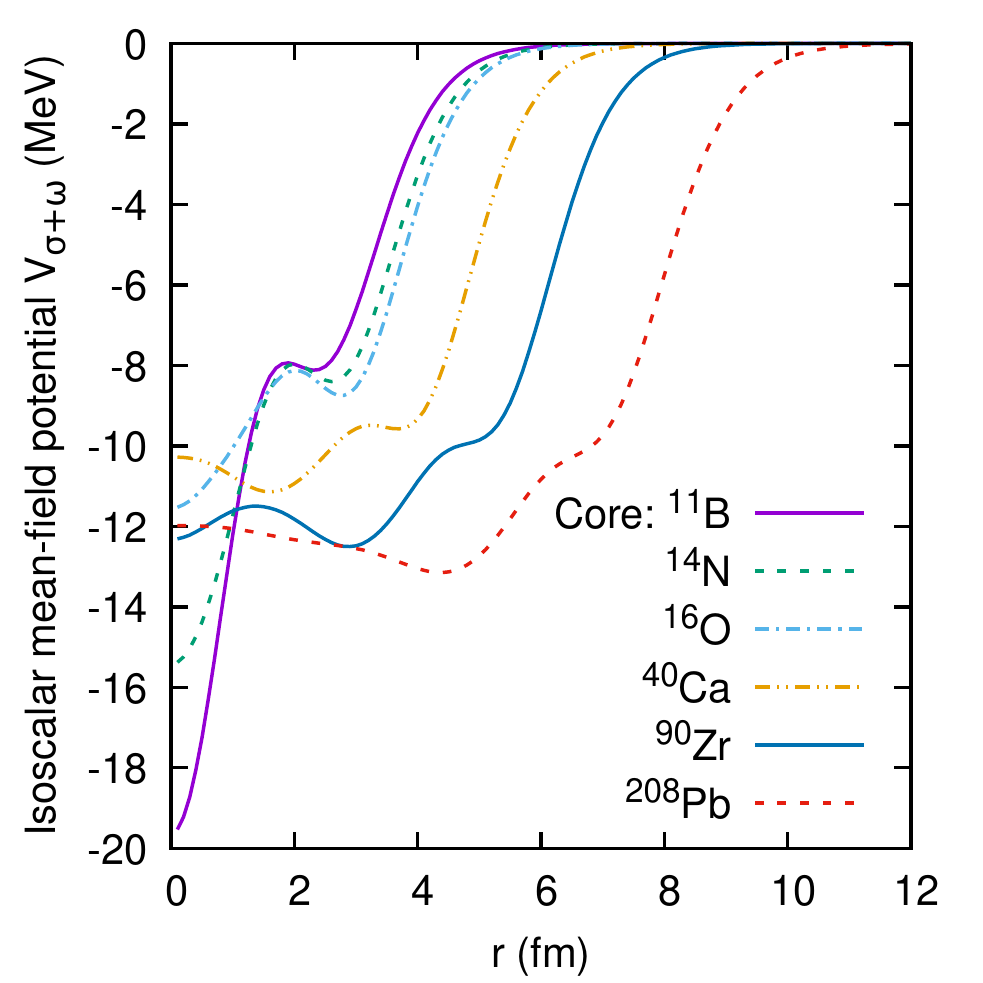}
\caption{The isoscalar part $V_{\sigma+\omega}$ of the mean field potential 
for $\Xi$ particle in light to heavy hypernuclei with $A=11-208$ cores. }
\label{fig:XNpot_sys}
\end{figure}

With the parameters thus obtained, we discuss  the isoscalar part of the $\Xi$ mean field potential, 
\begin{eqnarray}
V_{\sigma+\omega} = g_{\sigma\Xi}\sigma + g_{\omega\Xi}\omega^0, 
\end{eqnarray}
for various single-$\Xi$ hypernuclei with mass $A=12$ to $209$ shown in 
Fig. \ref{fig:XNpot_sys}.  
The potential increases smoothly from the center to the surface without a plateau in the interior  part of $\Xi$ potential for the  mass  $A \leq 20$.  The central depth is $-20$ MeV for $^{11}$B core and gradually become shallower for $^{14}$N and $^{16}$O cores.   
For heavier $\Xi$ hypernuclei with $A \geq 40$, the central part of the potential becomes flat and the depth  converges to be 
around $-12$ MeV in heavier systems, which can be considered as an infinite matter limit. 
The width of the potential increases to follow $A^{1/3}$ law.
 
\begin{figure}
\includegraphics[width=\linewidth]{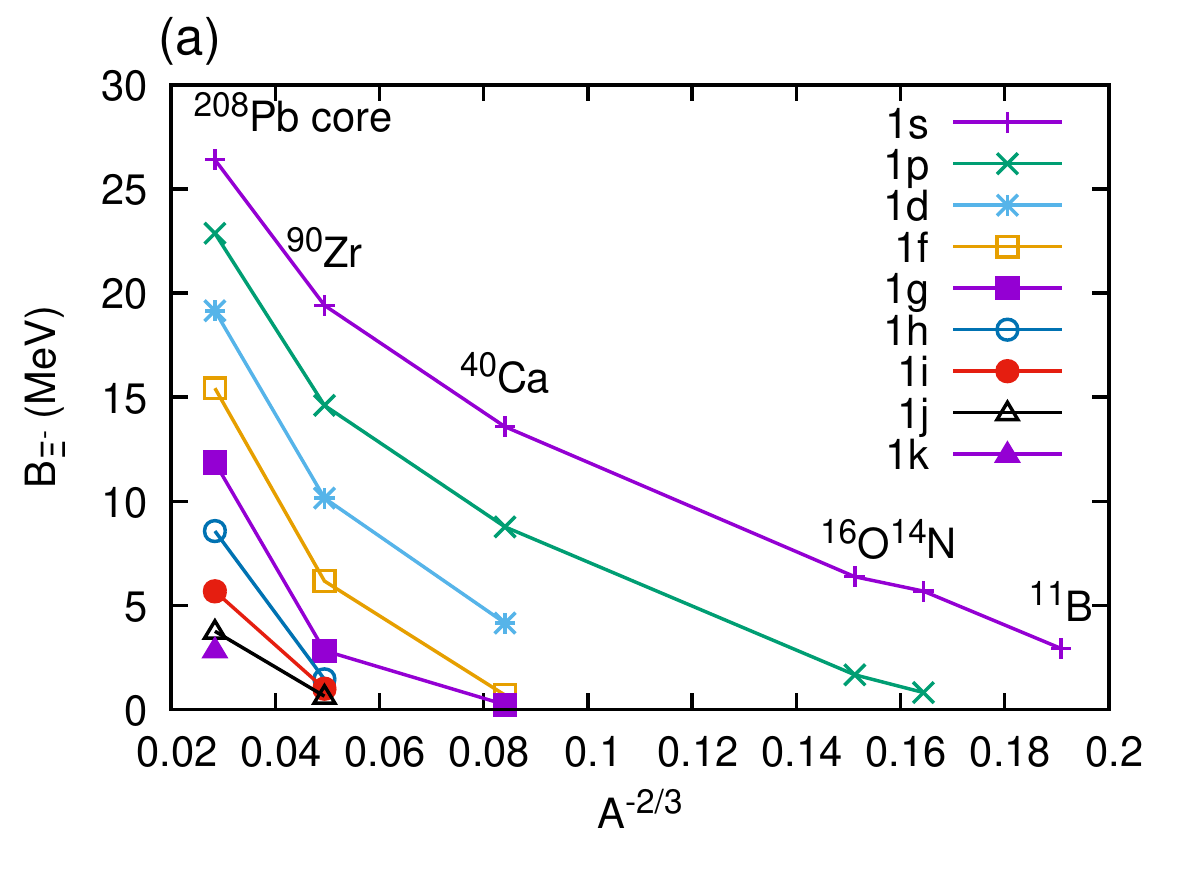}
\includegraphics[width=\linewidth]{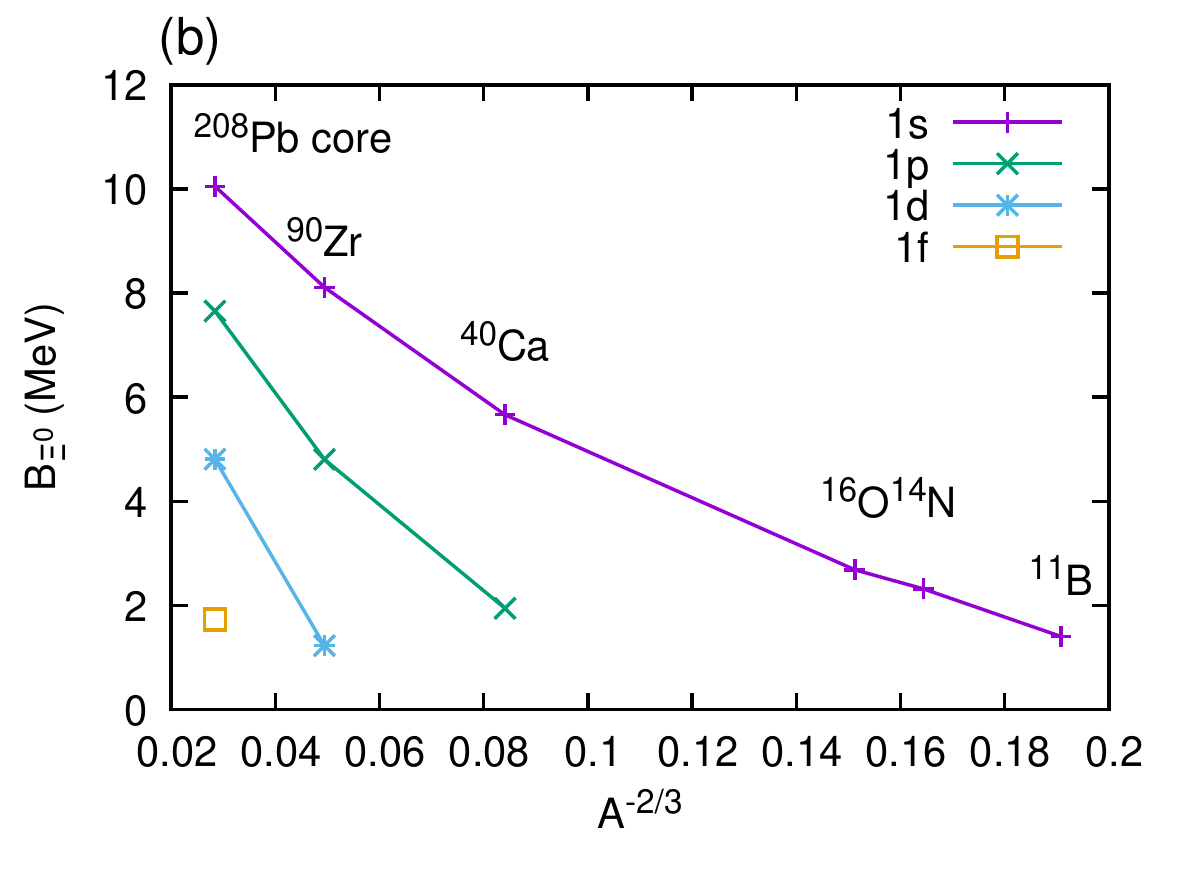}
\caption{Binding energies of single $\Xi$ states: (a) for $\Xi^-$ particle and (b) for $\Xi^0$ particle
in light to heavy hypernuclei with the mass $A=11-208$ core. The difference between (a) and (b) is with and without  Coulomb interaction between $\Xi$ particle and protons.}
\label{fig:BXi-_sys}
\end{figure}

The single-particle states of  $\Xi$ hypernuclei with mass $A=12$ to 209 are 
shown in Fig. \ref{fig:BXi-_sys}: (a) for $\Xi^-$ with the Coulomb interaction between a $\Xi^-$ particle and core protons; (b) for $\Xi^0$ without the Coulomb interaction.  
The Coulomb interaction between $\Xi^-$ and protons are attractive and makes the potential depth deeper about $-17$ MeV for $^{208}$Pb core and $-2$ MeV in $^{11}$B core.  In Fig. \ref{fig:BXi-_sys}(a), the $s$-orbit bounds $B_\Xi=27$ MeV for $^{209}_\Xi$Tl, and a light 
$\Xi$ hypernucleus $^{12}_\Xi$Be has a small $B_\Xi= 3$ MeV. 
In $^{209}_\Xi$Tl, the orbits only up to $l=8$ are shown although there a lot more 
Coulomb-assisted bound states. 
Only the $s$-orbit is bound in the $\Xi$ mean field potential of $^{12}_\Xi$Be.  
Due to the contribution of Coulomb interaction,
the behaviour of the single particle energies of $\Xi^-$ hypernuclei is
similar to that of $\Lambda$ hypernuclei \cite{TH12}.

In Fig. \ref{fig:BXi-_sys}(b), the $s$-orbit is bound in all $\Xi^0$ nuclei starting from $^{11}$B core to $^{208}$Pb core.   The binding energy of the   
 $s$-orbit 
of $^{208}$Pb core is about 10 MeV, which is almost the same as the
potential depth. 
For $A=11,\ 14$ and $16$ cores, only the $s$-orbit is bound, but 
the $p$-orbit of $\Xi^0$ hypernucleus becomes bound for  $A \geq 40$ core. 
For $^{208}$Pb core, the states with the angular momentum  $l \leq 3$ are bound. 

\section{Summary}\label{sec:summary}

We studied the energy spectra of $\Xi$ hypernuclei $^{15}_{\Xi}$C and $^{12}_{\Xi}$Be with an RMF model taking into account 
meson exchange $N\Xi$ interactions.  The mean field potential parameters are optimized to reproduce the average energy  of KINKA and IRRAWADDY events for possible ground states and also the average energy of KISO and IBUKI events for possible excited states in $^{15}_{\Xi}$C. 
We further introduce the two-body $s$-wave interaction between valence nucleons and $\Xi$ particle.  The strength of $s$-wave interaction is extracted from HAL Lattice QCD calculations, which gives a strong attraction in the $(T,S)=(0,0)$ channel.  We demonstrate that the HAL $s$-wave interaction has rather small effect on the energy of $^{15}_{\Xi}$C since the core nucleus $^{14}$N has the isospin 
$T$=0 and the isoscalar spin-spin interaction is rather small. We then introduce $p$-wave $N\Xi$ interaction to fit the energy difference between IRRAWADDY and KINKA event,  and found that both repulsive and attractive spin-spin $p$-wave $N\Xi$ interaction gives a reasonable energy splitting with the coupling strength 
$v_{\sigma}^p=-$150 MeV fm$^5$ or 100  MeV fm$^5$, respectively.  The attractive $p$-wave gives $J^\pi=1/2^+, 3/2^+$ states for the ground state and the first excited state, respectively.  On the other hand, the repulsive one give a reversed spectrum, i.e., $J^\pi=3/2^+, 1/2^+$ states for the ground state and the first excited state, respectively. 
These $p$-wave coupling strength gives also a reasonable energy
for the  IBUKI event as an excited $\Xi p$-state.  

The energy spectrum of $^{12}_{\Xi}$Be is also studied in the same theoretical model. The HAL $s$-wave interaction in the 
spin-isospin channel is strongly attractive so that the $(J^\pi,T)=(2^-,0)$ state becomes the lowest ground state and 
the states with  $(J^\pi,T)=(1^-, 1),(2^-, 1)$ and $(1^-,0)$ become the excited states in order.  When the $p$-wave interaction is introduced, the attractive  $p$-wave one gives more attractive energy for $(J^\pi,T)=(2^-,0)$ state, while the repulsive interaction cancels largely with the $s$-wave contribution each other and makes $(J^\pi,T)=(1^-, 1)$ state to be the ground state. 

The mean field $N\Xi$ potential of the present RMF model gives the potential depth $-12$ MeV in the nuclear matter limit. 
This potential depth is one-fourth of nuclear mean field and a half of $N\Lambda$ mean field.  These features might give a critical influence on the structure, especially the mass radius relation and the inert core of neutron star.

\begin{acknowledgments}
We would like to thank to Prof. Nakazawa for helpful discussion of the present
work. 
This work was supported in part by JSPS KAKENHI of Grant Numbers JP19K03858, JP19K03861, JP20H00155, Grand-In-Aid for scientific Research on Innovative Areas,
18H05407 and 
and the National Natural Science Foundation of China (Grant No. U2032141). 
\end{acknowledgments}

\end{document}